\documentstyle[11pt,aaspp4]{article}
\begin{document}

\title{High Ratio of $\rm ^{44}Ti$/$\rm ^{56}Ni$ 
       in Cas A and Axisymmetric
%/Asymmetric 
Collapse-Driven Supernova
Explosion }

\author{Shigehiro Nagataki\altaffilmark{1}, Masa-aki
Hashimoto\altaffilmark{2}, Katsuhiko Sato\altaffilmark{1,3,5},
Shoichi Yamada\altaffilmark{1,4}, and Yuko S. Mochizuki\altaffilmark{5}}

\noindent
\altaffilmark{1}{Department of Physics, School of Science, the University
of Tokyo, 7-3-1 Hongo, Bunkyoku, Tokyo 113, Japan}\\
\altaffilmark{2}{Department of Physics, Faculty of Science,
Kyusyu University, Ropponmatsu, Fukuoka 810, Japan}\\
\altaffilmark{3}{Research Center for the Early Universe, School of
Science, the University of Tokyo, 7-3-1 Hongo, Bunkyoku, Tokyo 113, Japan} \\
\altaffilmark{4}{Max-Planck-Institute f\"ur Physik und Astrophysik
Karl-Schwarzschild Strasse 1, D-8046, Garching bei M\"unchen, Germany\\
\altaffilmark{5}{The Institute of Physical and Chemical Research (RIKEN),
Hirosawa 2-1, Wako, Saitama, Japan 351-01
}
}

% Notice that each of these authors has alternate affiliations, which
% are identified by the \altaffilmark after each name.  The actual alternate
% affiliation information is typeset in footnotes at the bottom of the
% first page, and the text itself is specified in \altaffiltext commands.
% There is a separate \altaffiltext for each alternate affiliation
% indicated above.

\begin{abstract}
The large abundance ratio of $\rm ^{44}Ti$/$\rm ^{56}Ni$ in Cas A is
puzzling. In fact, the ratio seems to be larger than the theoretical
constraint derived by Woosley $\&$ Hoffman (1991).
However, this constraint is obtained on the assumption that
the explosion is spherically symmetric,
%.
%On the other hand, 
%Hardy
whereas 
Cas A is famous for the asymmetric form of the remnant.
Recently, Nagataki et al. (1997) 
%performed 
calculated 
the explosive
nucleosynthesis 
%under 
%for 
%Hardy
of 
axisymmetrically deformed collapse-driven supernova.
They reported that the ratio of $\rm ^{44}Ti$/$\rm ^{56}Ni$ was
enhanced by the 
%strong 
stronger 
alpha-rich freezeout in the polar region.
In this paper, we apply 
%the 
these 
results 
%of Nagataki et al. (1997) 
to Cas A 
and examine whether 
%its 
%Hardy
this 
effect can explain the large amount of $\rm
^{44}Ti$ and the large ratio of $\rm ^{44}Ti$/$\rm ^{56}Ni$.
We demonstrate that the conventional spherically
symmetric explosion model can not explain the $^{44}$Ti mass produced in Cas A 
if its 
lifetime 
%Hardy
%half life 
is shorter than $\sim$ 80 years and the 
%Hardy
intervening 
space is
transparent to the gamma-ray line from the decay of $^{44}$Ti.
On the other hand, we show the axisymmetric explosion models can
solve the problem. 
%We propose that asymmetric explosion models
%will also have a possibility to solve the problem since the strong
%alpha-rich freezeout will occur in that case, too.
We 
%Hardy
%can 
expect the same effect 
%for 
%Hardy
from a 
three dimensionally asymmetric
explosion, since the stronger alpha-rich freezeout will 
%occur also 
%Hardy
also occur 
in 
that case in the region where the larger energy is deposited. 

\end{abstract}


%%%%%%%%%%%%%%%%%%%%%%%%%%%%%%%%
\section{Introduction} \label{intro}
%%%%%%%%%%%%%%%%%%%%%%%%%%%%%%%%

\indent

Cassiopeia A (Cas A) is a young supernova remnant
which is located relatively close to us ($2.9 \pm 0.1$ kpc,
\cite{braun87}). There is an old 
record that John Flamsteed observed this supernova phenomenon (1680
AD?) and the
new star had an apparent brightness of the 6th magnitude (\cite{ashworth80}).
The mass of the progenitor is estimated to be (15-30) $M_{\odot}$
(\cite{hurford96}), which 
%proves 
implies 
that Cas A is the remnant of collapse-driven
supernova explosion.

Recently, Cas A was observed by the COMPTEL telescope aboard the
Compton Gamma-Ray Observatory and the gamma-ray line (1.157 MeV) from
$\rm ^{44}Ti$ decay was detected at 
%a 
the 
significance level of $\sim 4
\sigma$. The measured line flux was (7.0$\pm$1.7)$\times
10^{-5}$ photons/cm$^2$/sec (\cite{iyudin94}).
However, 
%they reported afterwards that the 
%Hardy
this 
flux 
%is recalculated to be
%Hardy
measurement was revised to 
(4.8$\pm$0.9) $\times 10^{-5}$ photons/cm$^2$/sec by 
%a 
%Hardy
%the 
further
analysis (\cite{iyudin97}).
On the other hand, OSSE observations reported that the flux 
is (1.7$\pm$1.4)$\times 10^{-5}$ photons/cm$^2$/sec (\cite{the96}).
At the 1 $\sigma$ level, the revised COMPTEL observation and the OSSE
result are in agreement near
3.5$\times 10^{-5}$ photons/cm$^2$/sec.
The inferred amount of $\rm ^{44}Ti$ is larger than 1 $\times 10^{-4}
M_{\odot}$, although 
%it 
%Hardy
this estimate 
depends on the distance and the age of Cas A, and
especially, on the half-life of $\rm ^{44}Ti$ which is still very uncertain
(
%lifetime measurements for time scales of human life are apparently
%difficult
it is difficult to measure half-life of an element of the order of the
human life time; we also note that the half-life
equals to the lifetime$\times \rm log_e 2$, by way of precaution
). It is reported to 
%be 
fall in the range from
$39 \pm 4$ yr (\cite{meibner96}) to $66.6 \pm 1.6$ yr
(\cite{alburger90}). This range corresponds to (50--98) yr in the lifetime.

%Anyway, the total mass of $\rm ^{44}Ti$ in Cas A is considerably more
%abundant compared with a
%theoretical prediction (\cite{woosley95}; \cite{timmes96}) even if the
%lower limit of the mass 1$\times 10^{-4} M_{\odot}$ is adopted.
%Hardy
%Anyway, the 
The 
observed mass of $\rm ^{44}Ti$ in Cas A, 1$\times 10^{-4} M_{\odot}$, is 
%Hardy
%considerably
large compared with the theoretical prediction 
(\cite{woosley95}; \cite{timmes96}) even if we adopt the
lower limit of 
%Hardy
the 
COMPTEL observation. 
Moreover, this amount of $\rm
^{44}Ti$ 
%necessitate 
is accompanied by 
the ejection of at least 0.05 $M_{\odot}$ of $\rm
%^{56}Ni$ if we assume 
%Hardy
^{56}Ni$, if we assume that  
%a 
the 
theoretical 
%constraint 
prediction 
for the ratio of mass
%fractions 
%Hardy
fractions, 
$X(\rm ^{44}Ti)/ \it X(\rm ^{56}Ni) \leq 1.89 \times 10^{-3}$, 
(\cite{woosley91})
is correct
. 
%If this is true,
However, this much $\rm ^{56}Ni$ would have led to 
the peak absolute magnitude 
%of the supernova would have been of the order
of 
$\sim$
-4, 
%which 
and 
would have been 
much
brighter than the recorded 6th magnitude of
the apparent brightness. This is the puzzling abundance problem of Cas A.

Although the 
%effect 
possibility 
of extinction of $\sim$ 10 mag is proposed 
%for an explanation of 
to solve 
this problem (\cite{timmes96}), it is 
%impossible to say definitely
%that such an effect is 
yet to be 
confirmed observationally (\cite{predehl95}).
In the present paper, we propose another explanation, that is,
the effect of 
%axisymmetry/
asymmetry of the supernova explosion.

%Historically speaking, 
%Hardy
Historically, 
almost
all 
%calculations of 
explosive nucleosynthesis 
calculations
with the use of a large nuclear reaction network have
been performed 
%on the assumption that the explosion is
%spherical symmetric. 
for 
%Hardy
%the 
spherically symmetric explosion models.  
On the other hand, both the image of Cas A inferred from optical and
X-ray observations
and
the velocity distribution inferred from optial observation
show that the
remnant is not spherically symmetric. Particularly in optical images,
a remarkable 
%feature which is generally called the jet 
jet-like feature 
is present at
the east-side (e.g., \cite{fesen96}).
%If the 
The 
fact that 
%a 
no 
pulsar 
%is not 
has been 
found 
yet 
in Cas A 
%is taken into
%consideration, which proves 
might imply that 
the jet 
%is 
%was 
%not originated from the pulsar, 
%Hardy
did not originate from the pulsar, 
but 
%
%the jet may have been related to 
was caused by 
the core-collapse dynamics, that is, the
explosion itself would have been asymmetric (e.g., \cite{burrows95}).

Recently, Nagataki et al. (1997) 
%Hardy
%have 
calculated for the first time
explosive nucleosynthesis in
axisymmetrically deformed collapse-driven supernovae 
%with the use of 
using 
a 
large nuclear reaction network 
%we note that although Bazan and Arnett
%have also performed 2-dimensional calculations, they has calculated
%the chemical evolution during the steady stellar evolution, that is,
%before the supernova explosion 
(see 
\cite{bazan97} 
for 2D nucleosynthesis calculations in the presupernova stages, and
refer also to M\"{u}ller et al. (1991) for explosive
nucleosynthesis with the Rayleigh-Taylor instability).
%please add the refernce! Yamada
They found that the ratio $X(\rm ^{44}Ti)/ \it X(\rm
^{56}Ni)$ can 
%get 
%Hardy
be 
significantly larger than that of the spherical explosion.
%They explained this result as follow:
The reason is as follows:

Since $\rm ^{44}Ti$ is synthesized through the alpha-rich freezeout, high
entropy is needed for the synthesis of this nucleus.
The relation between the energy density and the temperature behind the
shock wave is approximately $E \cong a T^4$, where $a$ is the
radiation constant.
Since the axisymmetric explosion generates 
a 
stronger shock wave in the
polar direction, 
a 
higher temperature is 
%achieved 
reached 
%along 
in 
this direction,
%. As a result, 
resulting in much higher 
%very high 
entropy per baryon
%, which is never
%obtained from 
than that for 
the spherical explosion. 
%, is realized in the polar region.
It should be noted 
again 
that 
%the axisymmetric explosion can achieve a high ratio of
it is this stronger alpha-rich freezeout that yields the higher ratio
of 
$X(\rm ^{44}Ti)/ \it X(\rm ^{56}Ni)$ 
for the axisymmetric explosion 
%
%above 
beyond 
the constraint (\cite{woosley91}) for the ratio in the
spherical explosions.
%because of this strong alpha-rich freezeout.

In the present paper, we investigate whether 
%the results of Nagataki et al. (1997) 
the axisymmetric explosion mechanism 
can explain the mass of $\rm ^{44}Ti$ and the high
ratio of $X(\rm ^{44}Ti)/ \it X(\rm ^{56}Ni)$ 
observed 
in Cas A.
In section \ref{nucleosynthesis}, we
summarize the results of Nagataki et al. (1997).
Comparison with 
%the observation 
%Hardy
observations 
is presented in section
\ref{results}. Summary and discussion are given in section \ref{summary}.

%%%%%%%%%%%%%%%%%%%%%%%%%%%%%%%%%%%%%%%%%%%%%%%%%%%%%%%%%%%%%%%%
\section{ Explosive Nucleosynthesis 
%under 
for 
Axisymmetric Supernova
%Explosion} \label{nucleosynthesis}
%Hardy
Explosions} \label{nucleosynthesis}
%%%%%%%%%%%%%%%%%%%%%%%%%%%%%%%%%%%%%%%%%%%%%%%%%%%%%%%%%%%%%%%%

\indent

\begin{figure}
\epsscale{1.0}
\plotone{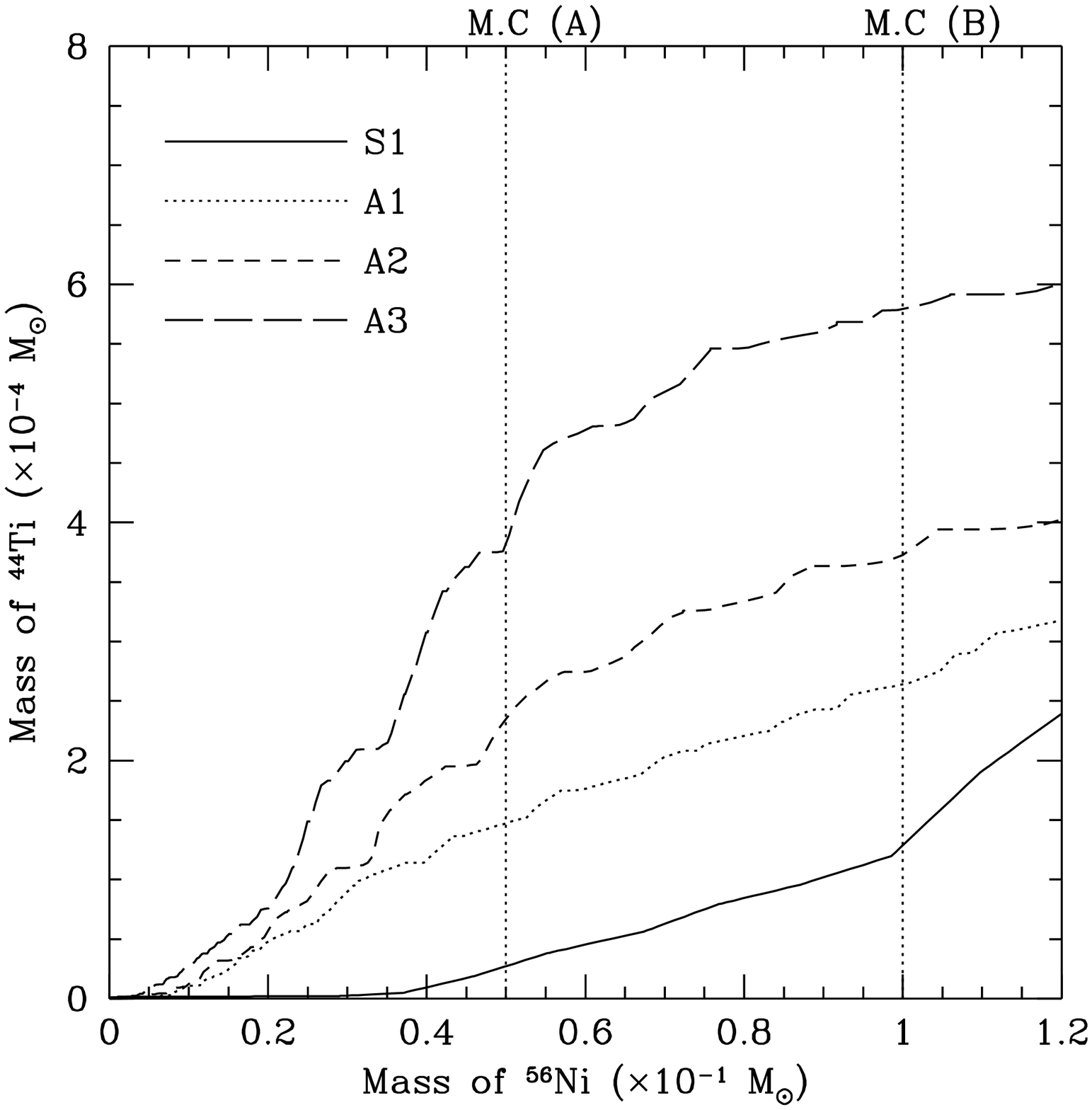}
%\figcaption{
%Relations between the mass of $\rm ^{56}Ni$ and that of
%$\rm ^{44}Ti$ with four models S1, A1, A2, and A3.
%Model S1 denotes spherical explosion, A1, A2, and A3 are
%axisymmetric explosions with increasing order of its degree.
%M.C (A) and M.C (B) indicate the amounts of $\rm ^{56}Ni$ which
%determine the mass cuts.
%\label{ti44ni56}}
%%%%%%%%%%%%Yamada: I would modify the figure caption as follows.
\figcaption{
Relations between the mass of $\rm ^{56}Ni$ and that of
$\rm ^{44}Ti$ with four models S1, A1, A2, and A3.
Model S1 denotes spherical explosion, A1, A2, and A3 are
axisymmetric explosions in increasing order of its degree of asymmetry.
M.C. (A) and M.C. (B) indicate the location of 
the mass cuts.
\label{ti44ni56}}
\end{figure}

In this section, we summarize the results of Nagataki et al. (1997).
They performed 2-dimensional hydrodynamical calculations and
%calculated 
studied 
the changes of the chemical
compositions 
%with the use of 
using 
%the 
%Hardy
a 
large 
nuclear reaction network containing
242 nuclear species.
%As for the mass cut, the 
The 
location 
of the mass cut 
was not determined from the hydrodynamical
calculation but from the amount of $\rm ^{56}Ni$ in the ejecta.
We note that all material inside the mass
cut are assumed to fall back 
%to 
onto 
the central compact object (this is
the definition of 
%a 
the 
mass cut).
For example, the mass cut 
for the model of SN 1987A
was determined so as to contain $0.07
M_{\odot}$ of $\rm ^{56}Ni$ in the ejecta
% for the model of SN 1987A.
, which is the value inferred from the light curve. 
We show in Figure~\ref{ti44ni56}
the relations between the ejected mass of $\rm ^{56}Ni$ and that of
$\rm ^{44}Ti$ for four models S1, A1, A2, and A3.
Model S1 
%denotes 
is a 
spherical explosion, A1, A2, and A3 are models of
axisymmetric explosions 
%with 
in 
an increasing order of the degree of
%axisymmetry.
asymmetry. 
For the axisymmetric models A1-3 the initial velocity behind the shock 
wave is assumed to be radial and proportional to $r \times 
\frac{1 + \alpha \cos (2 \theta)}{1 + \alpha }$, where $r$, $ \theta $, 
and $ \alpha $ are the radius, the zenith angle and the model
parameter that determines the degree of asymmetry, respectively. In
this study, we take $ \alpha = 0 $ for model S1, $ \alpha =
\frac{1}{3} $ for model A1, $ \alpha = \frac{3}{5} $ for model A2, and
$ \alpha = \frac{7}{9} $ for model A3. The larger $ \alpha $ gets, the more
asymmetric the explosion becomes. We assumed the same distribution
also for the thermal energy. 
%The half 
%Hardy
Half 
of the total energy 
%is given by 
%Hardy
appears as 
%Hardy
%the 
kinetic energy and the 
%rest 
%Hardy
other 
half 
%by the 
%Hardy
as 
thermal energy. 
Once the position of the mass cut is determined, the mass of $\rm
^{44}Ti$ is also obtained for each model as is seen from
Figure~\ref{ti44ni56}.

\placefigure{ti44ni56}

%%%%%%%%%%%%%%%%%%%%%%%%%%%%%%%%%%%%%%%%%%%%%%%%%%%%%%
%\section{Comparison with the observation} \label{results} 
%Hardy
\section{Comparison with observations} \label{results} 
%%%%%%%%%%%%%%%%%%%%%%%%%%%%%%%%%%%%%%%%%%%%%%%%%%%%%%
\indent

The amount of $\rm ^{44}Ti$ synthesized in Cas A is estimated from the 
observed flux by using  the
following equation 
%Hardy
which holds 
if the 
%space 
intervening matter 
is transparent to the gamma-ray line:
\begin{eqnarray}
M_{\odot}(\rm ^{44}Ti) = 1.38 \times 10^{-4}\frac{F_{\gamma}}{1 \rm
cm^{-2}s^{-1}}\left(\frac{\it d}{\rm 1kpc}\right)^2
%\left(\frac{\itM(\rm ^{44}Ti)}{44\it m_u}\right)
\left(\frac{\tau}{\rm 1yr}\right)
%\exp(-t/\tau)
\exp(t/\tau) \quad,
\label{eqn1}
\end{eqnarray}
where $M_{\odot}(\rm ^{44}Ti)$, $F_{\gamma}$, $d$, $M(\rm ^{44}Ti)$, 
%$m_u$,
$\tau$, and $t$ are the amount of $\rm ^{44}Ti$ (in solar mass) 
synthesized in Cas A, 
the gamma-ray flux, the distance, the mass of $\rm
^{44}Ti$, 
%the atomic mass unit, 
the 
lifetime 
%Hardy
%half life 
of $\rm ^{44}Ti$, and the 
age of Cas A, respectively.

\begin{figure}
\epsscale{1.0}
\plotone{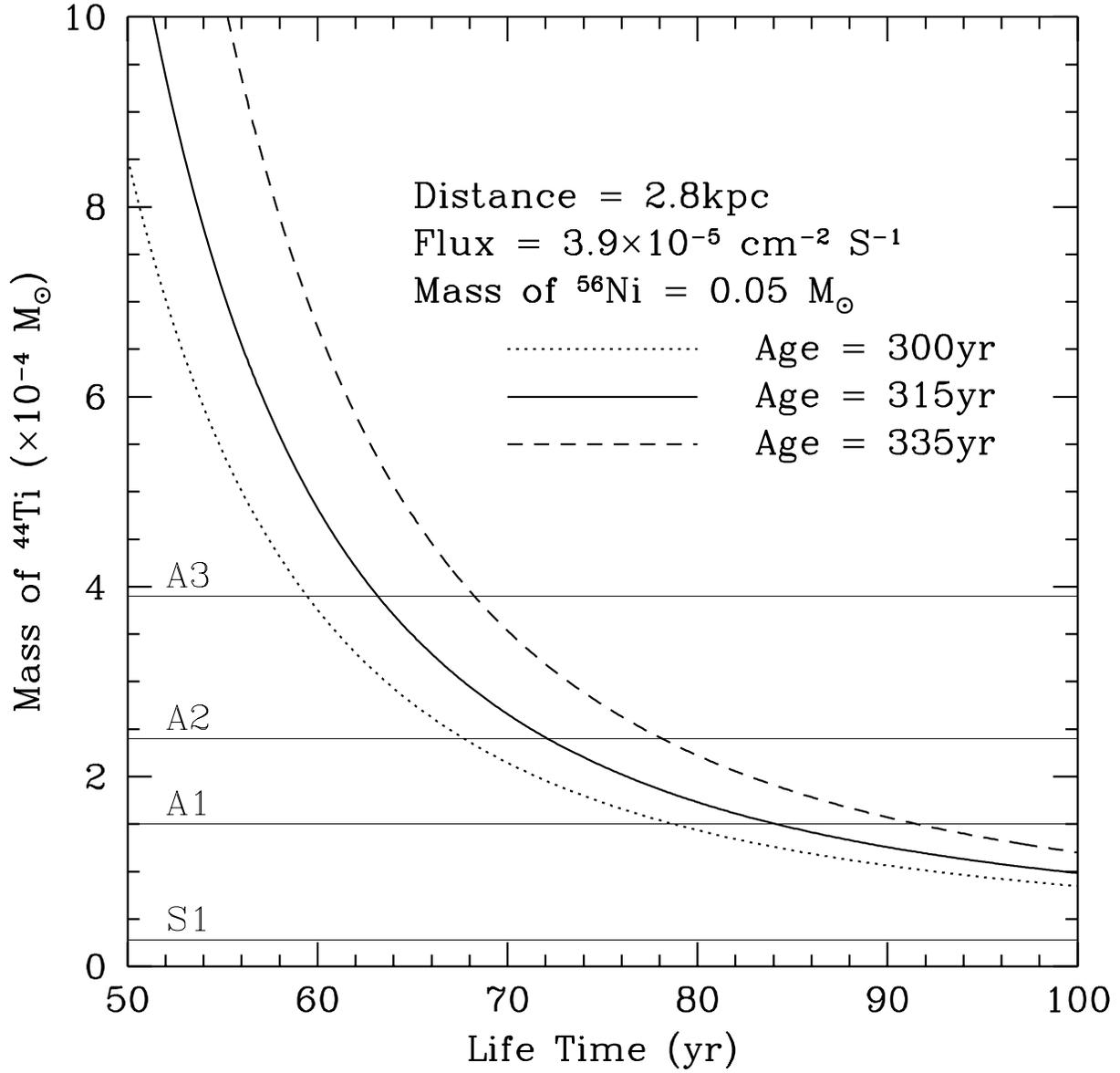}
\figcaption{
Relation between the amount and the lifetime of $\rm ^{44}Ti$
for the lowest flux (3.9 $\times \rm 10^{-5} cm^{-2}s^{-1}$) and the
nearest distance
(2.8 kpc). The age of Cas A is assumed to be 300, 315, and 335 yr.
Horizontal lines mean the ejected amount of $\rm ^{44}Ti$ for the
four models.
The ejected mass of $\rm ^{56}Ni$ is set to be $0.05 M_{\odot}$ (this
corresponds to M.C (A) in Figure 1).
\label{fig2}}
\end{figure}

\begin{figure}
\epsscale{1.0}
\plotone{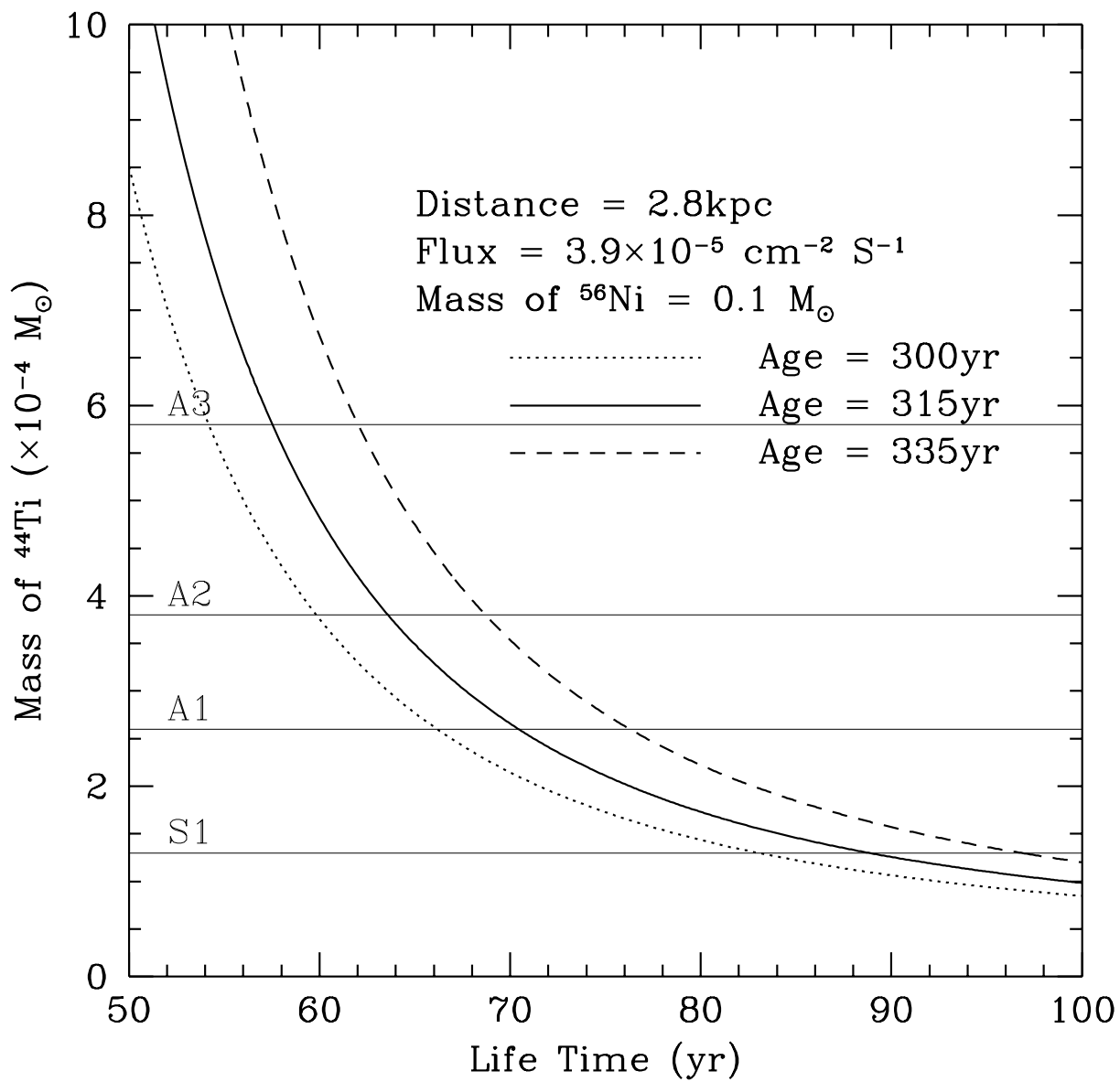}
\figcaption{
Same as Figure 2 but for the amount of $\rm ^{56}Ni =
0.1 \it M_{\odot}$ (this corresponds to M.C (B) in Figure 1).
\label{fig3}}
\end{figure}

We show in Figures ~\ref{fig2} and~\ref{fig3} the amount of $\rm
^{44}Ti$ as a function of the lifetime. Since the age of Cas A is
uncertain, we study three cases; 300, 315, and 335 yrs (\cite{vandenbergh83}).
We set other parameters 
so as 
to be modest for the spherical explosion 
model.
%
%We will see that spherical explosion can hardly
%explain the amount of $\rm ^{44}Ti$ even with these modest parameters.
As one can see from the equation~\ref{eqn1}, the lowest amount of $\rm
^{44}Ti$
allowed by the observation is obtained by choosing 
the smallest values for the flux and the distance within
%Hardy
the 
uncertainties. 
Therefore, we adopt in this paper 
the lowest flux of the COMPTEL telescope (3.9 $ \times \rm 10^{-5}
cm^{-2}s^{-1}$) and the nearest distance (2.8 kpc). 
We will see that the spherical explosion model can 
%hardly
%Hardy
not 
explain the amount of $\rm ^{44}Ti$ even with these modest parameters.

For comparison, we give the amount of $\rm ^{44}Ti$ ejected from the
four models. 
Since 
%
%Unlike SN 1987A, 
the amount of $\rm ^{56}Ni$ in Cas A is unknown,
%
%on the contrary to SN 1987A, 
%Hardy
unlike SN 1987A, 
%. Because of this reason, 
we investigate the mass of $\rm ^{44}Ti$ 
%with
for 
two mass cuts (M.C (A) and M.C (B) in Figure~\ref{ti44ni56}) to see its effect.
In 
%one case,
case of M.C.(A), 
$0.05 M_{\odot}$ of $\rm ^{56}Ni$ is 
%assumed to be 
ejected
and $0.1 M_{\odot}$ in 
%the other
M.C.(B). The results are 
%drawn 
shown 
in
Figures ~\ref{fig2} and ~\ref{fig3} by the horizontal lines.

We can see from these figures the spherical explosion can 
%hardly
%Hardy
not 
produce enough amount of $\rm ^{44}Ti$ even if the youngest age,
the nearest distance 
within uncertainty, 
and a 
%much 
rather larger 
amount of $\rm ^{56}Ni$ are used.
Therefore, we conclude that the spherically symmetric explosion model 
%fails in 
is unlikely to 
%explaining 
explain 
the $^{44}$Ti mass 
%produced 
observed 
in Cas A 
if the lifetime is shorter than $\sim$ 80 years and the 
%space 
intervening matter 
is
transparent to the gamma-ray line.
On the other hand, 
it is shown 
the axisymmetric 
%model 
models 
can produce 
%much 
larger 
amount of
$\rm ^{44}Ti$
%beyond the upper limit in 
than 
the spherical explosion.
It should be noted that no spherically symmetric explosion models
%proposed 
calculated 
so far
predict more than $\sim 1.5 \times 10^{-4} M_{\odot}$ of $^{44}$Ti
in agreement with our model S1 
(\cite{woosley95}; \cite{hashimoto95}).

\placefigure{fig2}
\placefigure{fig3}

%%%%%%%%%%%%%%%%%%%%%%%%%%%%%%%%%%%%%%%%%%%%%%%%%%%%%%%%%%
\section{ Summary and Discussion} \label{summary}
%%%%%%%%%%%%%%%%%%%%%%%%%%%%%%%%%%%%%%%%%%%%%%%%%%%%%%%%%%

\indent

We have tried to explain the observation of the large amount of $\rm
^{44}Ti$ and the high ratio of $\rm ^{44}Ti/
^{56}Ni$ in Cas A using the results of the axisymmetric explosive
nucleosynthesis by Nagataki et al. (1997).
We have found that the spherically symmetric explosion model 
%fails in 
is unlikely to give an explanation 
%to
%Hardy
of  
%explaining 
such a high ratio 
of $\rm ^{44}Ti / ^{56}Ni$
if the lifetime of $\rm ^{44}Ti$ is shorter than $\sim$ 80 years and
the 
%space 
intervening matter 
is
transparent to the gamma-ray line from the decay of $^{44}$Ti.
%On the other hand, we have shown that the axisymmetric models can solve
%the problem.
%
%
Although 
%the 
%Hardy
a 
large extinction of $\sim$ 10 mag due to the intervening 
matter was proposed as a solution to the problem of
%the darkness of the 
the observed low 
apparent brightness (\cite{timmes96}), 
%it is 
%impossible to say definitely that such an effect is 
%remaining 
%Hardy
this remainds 
to be 
confirmed observationally 
%so far 
(\cite{predehl95}).
%At least, as 
Instead, we 
%paid our attention to 
%Hardy
based our model on 
the observational fact that  
the shape of Cas A is far from spherically symmetric 
%from the observation 
(e.g., \cite{fesen96}),
%
%, such an effect should be taken into consideration for the calculation of the
%explosive nucleosynthesis.
%We believe our proposal of the effect of the axisymmetric explosion is
%worth while being stressed.
and we found by 2D calculations that the axisymmetric explosion models 
can also solve this problem. 
We think it is qualitatively correct that the ratio of $\rm ^{44}Ti /
^{56}Ni$ is enhanced by axisymmetric explosion, since larger energy is 
deposited near the rotational axis and higher entropy per baryon
is reached there, resulting in stronger alpha-rich freezeout and
yielding the larger amount of $\rm ^{44}Ti$ and higher ratio 
of $\rm ^{44}Ti / ^{56}Ni$.  Quantitatively, however, there are
uncertainties in the location of mass cut as well as in the initial degree of
asymmetry, both of which are difficult to 
%get 
%Hardy
derive 
from analytical
%studies or numerical simulations at least at present. In spite of 
%Hardy
studies or numerical simulations, at least at present. In spite of 
these quantitative difficulties, we believe 
%our scenario is
%Hardy
it is worthwhile for our scenario to be studied further,
%worthwhile to be studied further, 
since 
%it turned out that it could be the 
%Hardy
it results in a 
solution to the Cas A problem. 

%As for the mass of the progenitor,
%Nagataki et al. (1997) calculated explosive nucleosynthesis with a 6
%$M_{\odot}$ He core model 
In this paper we used only one progenitor model, where 
%Hardy
the 
He core mass is 
6 $M_{\odot}$, 
which corresponds to 18-21 $M_{\odot}$ in the
main-sequence stage. Since the progenitor of Cas A is thought to be
15--30 $M_{\odot}$, 
%it 
%Hardy
one 
may need to examine the dependence of our
conclusion on the progenitor mass. However, we think the
qualitative tendency for the ratio of $\rm ^{44}Ti/ ^{56}Ni$
will not be changed
for the following reason. 
The entropy per
baryon after the passage of the shock wave is 
%proportional approximately 
%Hardy
approximately proportional 
to $T^3/ \rho$, where $T$ and $\rho$ are the local
temperature and density. This means the entropy
per baryon depends more strongly on the temperature 
%rather than the
%Hardy
than on the 
density. The temperature depends significantly on the strength
of the initial shock wave
%and 
while 
the density depends on the initial structure of the
progenitor. Since the iron cores of the presupernova models have
similar structures (\cite{hashimoto95}), the qualitative tendency for
the ratio of $\rm ^{44}Ti/^{56}Ni$ will be unchanged even if the mass
of the progenitor is changed.

%The essence of our result is 
The essential point of our results is that 
%not that the form of the
%explosion is axisymmetric but 
%that the explosion is not spherically
%symmetric, which
%means 
the explosion energy is localized in 
%some small regions
a small region near the rotation axis
%. As a result, 
, leading to the stronger alpha-rich freezeout there.
%the strong alpha-rich freezeout occurs.
%Therefore, we can expect a large amount of 
Therefore, we expect 
%Hardy
that 
the same mechanism 
%also works 
%Hardy
will also work 
for 
three dimensionally asymmetric explosions, where the explosion energy
is deposited in some small regions. 
%yield of 
%$\rm ^{44}Ti$ even if the 
%explosion is not axisymmetric but 
%
%three dimensionally 
%
%asymmetric.

There are 
%two main effects 
%mainly two possibilities conceivable 
%Hardy
two main possibilities conceivable 
that make the shock wave
%axisymmetric/
asymmetric in 
the 
collapse-driven supernova explosion.
One is the effect of rotation (and/or magnetic field) of the
progenitor, which causes the shock wave 
%axisymmetric
%Hardy
axisymmetry 
(\cite{muller81}; \cite{symbalisty84}; \cite{mochmeyer89}; \cite{yamada94}). 
%Please add more references here! Yamada 
The other is the effect of 
neutrino-driven convection (\cite{shimizu94}; \cite{burrows95}) which
causes high-speed ``fingers'' in the mantle
%. In that case, an 
, resulting in a more complex 
asymmetric explosion.
% may occur in a more complex way. 

Although 
%results of 
in this paper and 
in Nagataki et al. (1997), 
we 
 have shown the effect of the 
strong alpha-rich freezeout for the 
2D 
axisymmetric explosion,
we also want to stress the necessity of 
%calculating 
explosive nucleosynthesis
calculations
with a large nuclear reaction network
for the ``finger-like'' explosion. 
Although it requires very high mesh resolution and CPU-time, we think
such a calculation will be
necessary to explain the 
observed complex structure
%phenomena 
of Cas A
% more quantitatively.
. 
We 
%will perform 
are planning 
such a calculation in the near future.

\acknowledgements
We would like to thank an anonymous referee for useful comments.
We are also grateful to S. Hardy and L. van Waerbeke for their
kind review of the manuscript.
This research has been
supported in part by a Grant-in-Aid for the Center-of-Excellence (COE) 
Research (07CE2002) and for the Scientific Research Fund (05243103,
07640386, 3730) of the Ministry of Education, Science, and Culture in
Japan, 
and by 
%Hardy
the 
Japan Society for the Promotion of Science Postdoctoral
Fellowships for Research Abroad.

\vskip1.0cm

\end{document}